\documentclass[11pt,fleqn]{article}

\usepackage{amsfonts,amssymb,cite}
\usepackage{epsf,graphicx}

\newcommand{\bls}[1]{\renewcommand{\baselinestretch}{#1}}

\def\noi{\noindent}

\newcommand{\Title}[1]{\noi {{\Large\bf #1}}\\[1ex]}

\newcommand{\Author}[2]{\noi{\bf #1}\\[2ex]\noi{\normalsize\it #2}\\}

\newcommand{\Abstract}[1]{\vskip 2mm \begin{center}
        \parbox{16.4cm}{\small\noi #1} \end{center}\medskip}

\newcommand{\foom}[1]{\protect\footnotemark[#1]}

\def\email#1#2{\footnotetext[#1]{e-mail: #2}\addtocounter{footnote}{1}}

\def\nq{\hspace*{-1em}}
\def\nqq{\hspace*{-2em}}

\def\cm{\hspace*{1cm}}
\def\inch{\hspace*{1in}}

\def\Jl#1#2{#1 {\bf #2},\ }

\def\ApJ#1 {\Jl{Astroph. J.}{#1}}
\def\CQG#1 {\Jl{Class. Quantum Grav.}{#1}}
\def\DAN#1 {\Jl{Dokl. AN SSSR}{#1}}
\def\GC#1 {\Jl{Grav. Cosmol.}{#1}}
\def\GRG#1 {\Jl{Gen. Rel. Grav.}{#1}}
\def\JETF#1 {\Jl{Zh. Eksp. Teor. Fiz.}{#1}}
\def\JETP#1 {\Jl{Sov. Phys. JETP}{#1}}
\def\JHEP#1 {\Jl{JHEP}{#1}}
\def\JMP#1 {\Jl{J. Math. Phys.}{#1}}
\def\NPB#1 {\Jl{Nucl. Phys. B}{#1}}
\def\NP#1 {\Jl{Nucl. Phys.}{#1}}
\def\PLA#1 {\Jl{Phys. Lett. A}{#1}}
\def\PLB#1 {\Jl{Phys. Lett. B}{#1}}
\def\PRD#1 {\Jl{Phys. Rev. D}{#1}}
\def\PRL#1 {\Jl{Phys. Rev. Lett.}{#1}}

\def\lal{&&\nqq {}}
\def\eq{Eq.\,}
\def\eqs{Eqs.\,}
\def\beq{\begin{equation}}
\def\eeq{\end{equation}}
\def\bear{\begin{eqnarray}}
\def\bearr{\begin{eqnarray} \lal}
\def\ear{\end{eqnarray}}
\def\earn{\nonumber \end{eqnarray}}

\def\nnn{\nonumber\\ \lal }

\def\yy{\\[5pt] {}}
\def\yyy{\\[5pt] \lal }

\def\dst{\displaystyle}

\def\eps{\varepsilon}

\usepackage{graphicx}
\bls{1.05}
\begin{document}
\twocolumn[

\Title{The conic-gearing image of a complex number \yy
    and a spinor-born surface geometry}

\Author{Alexander P. Yefremov\foom 1}
  {Institute of Gravitation and Cosmology of Peoples' Friendship University
  of Russia \\
  6 Miklukho-Maklaya St., Moscow 117198, Russia}

\Abstract
 {Quaternion (Q-) mathematics formally contains many fragments of physical
  laws; in particular, the Hamiltonian for the Pauli equation automatically
  emerges in a space with Q-metric. The eigenfunction method shows that any
  Q-unit has an interior structure consisting of spinor functions; this
  helps us to represent any complex number in an orthogonal form associated
  with a novel geometric image (the conic-gearing picture).
  Fundamental Q-unit-spinor relations are found, revealing the geometric
  meaning of spinors as Lam\'e coefficients (dyads) locally coupling the base
  and tangent surfaces.  }

\bigskip

] 
\email 1 {a.yefremov@rudn.ru}

\section{Introduction}

  Among hardly conceived physical objects, like the electric charge or the
  elementary particle spin, the most mysterious champion is probably the
  quantum-mechanical wave function. Providing extremely precise predictions
  of real phenomena, the theory of quantum mechanics remains in many aspects
  a ``given-by-heavens'' perfect computing technology with a ``noumenon''
  essence. Difficulties to reveal the essence have eventually overwhelmed
  attempts to endow the wave function with a geometrically conceivable
  image, and the majority of physicists of the 20th century conceded at its
  interpretation as a probability amplitude. A similar lack of understanding
  exists with spinor functions, relativistic extensions of the wave function,
  in the theory of Dirac or Weyl fermion fields. Indeed, human mind has
  usually no trouble in creating visual models of scalar, vector and maybe
  some tensor objects, but ``a square root from a vector'' (which a spinor
  mathematically is) makes the imagination fail.

  A search for a better perception of physical laws follows nowadays some
  new ways; one of them is a thorough investigation of hypercomplex (H-)
  numbers. Among the latest examples one can cite the ``perplex''
  (split-complex, double) numbers used to formulate 2-dimensional
  Lorentz-Minkowski relations [1] and the bi-quaternion (BQ-) numbers whose
  algebra is noticed to contain the basic formulas of relativity theory as
  its fundamental part [2, 3]. A closer view reveals at least two advantages
  of the H-number approach, especially that of quaternions. First, these
  numbers, the from times of Hamilton, their inventor [4], are recognized
  to be ``very geometric,'' giving clear vector images for the physical
  objects they are attributed to. Second, being represented by square
  matrices the numbers are found to be closely linked to the spinors.

  But there is one more advantage. Investigation of the whole set of
  H-numbers helps one to think of the formerly well-known simpler numbers as
  of a particular case of the set, and this may lead to surprising results.
  Representatives the of algebra of complex (C-) numbers are such
  doubtlessly known simpler numbers with their famous geometrical images on
  Wessel's complex plane or on Riemann's sphere. Nonetheless, it is shown
  below that a C-number regarded as a special quaternion, apart from its
  ``old'' properties, possesses some new features including an original
  geometric image. Written in the matrix form, the C-number is proved to
  inevitably have an interior structure whose elements are spinor functions.
  The functions further on are discovered to behave as dyad coefficients
  linking differentials of coordinates to a couple of specific
  2-dimensional (2D) surfaces.

  In Section II, a short review of H-number sets is given with necessary
  remarks on the stability of the numbers' multiplication table. In Section
  III, a C-number is treated as a dimensionally reduced quaternion in a
  matrix form, and its structural elements are investigated in detail.
  Section IV is devoted to a geometric interpretation of the C-number, its
  basic elements, and their graphic realization. Section V deals with a
  comparison of properties of the new geometry with those of spinors.
  The compact Section VI concludes the study.

\section{Units of quaternion numbers and the Pauli spin term}

  Frobenius' famous theorem states that the algebra of Hamilton's
  quaternions is the largest one in the number of basic units (four units
  with real coefficients) that admits a non-communicative but still
  associative multiplication (over addition). The next, by the number of
  dimensions, algebra of octonions (Kaylay's algebra, eight units) has
  non-associative multiplication. The most extended associative algebra is
  that of BQ-numbers built on Q-units, but with C-number coefficients at
  them. This algebra is not a ``good one'' since the norm of its numbers is
  not in general definable, and moreover, it contains zero-devisors. The
  BQ-algebra encompasses all other associative algebras: those of Q-numbers,
  C- and real numbers (good algebras), and two ``exotic'' algebras of
  split-complex and dual numbers, both with zero-devisors [5]. Let us recall
  here the main properties of Q-units, the base of all these sets.

  The BQ- and Q- algebras are based on one real (scalar) and three imaginary
  (vector) units \{$1; {\bf i, j, k}$\} with the following 16 equalities
  of the multiplication table postulated by Hamilton:
\bearr
     1^2  =-{\bf i}^2  =-{\bf j}^2  =-{\bf k}^2  =1,
\nnn
        {\bf i}1=1{\bf i}={\bf i},\ \ \
    {\bf j}1=1{\bf j}={\bf j},\ \ \
    {\bf k}1=1{\bf k}={\bf k},
\nnn                              \label{yef__1_}
    {\bf i}{\bf j}=-{\bf j}{\bf i}={\bf k},\ \ \
    {\bf j}{\bf k}=-{\bf k}{\bf j}={\bf i},\ \ \
    {\bf k}{\bf i}=-{\bf i}{\bf k}={\bf j}.
\ear
  If the units are written in the shorter (tensor) form
\bearr \label{yef__2_}
    {\bf i}={\bf q}_{1}, \ \ \  {\bf j}={\bf q}_{2} ,  \ \ \
    {\bf k}={\bf q}_{3} ,
\nnn
    \{1,  {\bf i,  j,  k}\}\mapsto \{ 1,  {\bf q}_{k} \},
\ear
  with small Latin indices $k,l,m,n...=1, 2, 3$, then the bulky set
  (\ref{yef__1_}) is reduced to
\beq \label{yef__3_}
    1 {\bf q}_{k} =  {\bf q}_{k}  1= {\bf q}_{k} ,\ \ \
    {\bf q}_{k} {\bf q}_{n} =-\delta_{kn} +\eps_{knm} {\bf q}_{m},
\eeq
  where $\delta_{kn} $ is the 3D Kronecker symbol, $\eps_{knm} $ is totally
  skew-symmetric 3D Levi-Civita symbol, and summing in repeated indices is
  implied. The multiplication table (\ref{yef__3_}) is convenient for an
  analysis of its stability (form-invariance) under transformations of
  Q-units; it turns out that there are two types of admissible
  transformations of the vector units, the scalar unit remaining
  intact [2]. The first type is the rotational (vector-type) transformation
\beq \label{yef__4_}
    {\bf q}_{k'} =  O_{k'n}  {\bf q}_{n}
\eeq
  where $ O_{k'n} $ is a $3\times 3$-matrix with components being in general
  C-numbers. After substitution of \eq (\ref{yef__4_}) into \eq
  (\ref{yef__3_}), the form of the multiplication rule (\ref{yef__3_}) is
  preserved if the matrix $ O_{k'n} $ satisfies the orthogonality condition
\beq \label{yef__5_}
    O_{k'n} O_{m'n} =\delta_{km} ;
\eeq
  this means that all transformations of the type (\ref{yef__4_}) form
  the special orthogonal group of 3D rotations over the field of C-numbers:
  $ O_{k'n} \in SO(3,C)$. The second type of transformations (reflections)
  that can give the same Q-triad (\ref{yef__4_}) is performed by the
  operator $ U $ and its inverse $ U^{-1} $:
\beq \label{yef__6_}
    {\bf q}_{k'} =  U {\bf q}_{k} U^{-1} ;
\eeq
  one immediately verifies that the transformation (\ref{yef__6_}) does not
  change the form of the rule (\ref{yef__3_}). The operators $ U $ are known
  to form the group of special linear 2D transformations over the field of
  C-numbers, i.e., $ U \in SL(2,C)$ (the spinor group), and this spinor
  group is 2:1 isomorphic to $ SO(3,C)$ and similarly to the Lorentz group.
  Thus the Q- (and BQ-) basis (\ref{yef__2_}), a pure algebraic object,
  turns out to be tightly linked to geometry and physics: vectors,
  rotations, spinors, and transformations of special relativity.

  The Q-units satisfying the rules (\ref{yef__3_}) can be described in terms
  of real and complex numbers, but these numbers should be components of
  square matrices. A simple $2\times 2$ matrix representation had been (up to
  a constant factor) proposed by Pauli, this representation is considered
  to be canonical:
\bearr \label{yef__7_}
     1= \left(\begin{array}{cc} {1} & {0} \\ {0} & {1} \end{array}\right),
     \qquad
    {\bf q}_{1} =-i \left(\begin{array}{cc} {0} & {1} \\
        {1} & {0} \end{array}\right),
\nnn
    {\bf q}_{2} =-i\left(\begin{array}{cc} {0} & {-i} \\ {i} & {0}
    \end{array}\right), \quad
    {\bf q}_{3} =-i \left(\begin{array}{cc} {1} & {0} \\ {0} & {-1}
        \end{array}\right);
\ear
  other samples can be obtained with the help of the transformations
  (\ref{yef__4_}) or (\ref{yef__6_}).

  Quaternions are in many ways linked to descriptions of physical laws, not
  only as a good tool but also as a ``math-medium'' where the physical
  relations known from experiment or formulated heuristically are
  surprisingly found. Apart from the above-mentioned connection with
  special relativity, a famous ``quaternion coincidence'' was revealed by
  Fueter [6] who discovered that the Cauchy-Riemann-like differentiability
  conditions for vector functions of a quaternion variable, pure
  mathematical requirements, are an exact formal equivalent of the full set
  of the vacuum Maxwell equations. Another striking example is the Pauli
  quantum-mechanical Hamiltonian computed in the Q-metric. Let an
  electron with mass $m$ and electric charge $e$ in an exterior magnetic
  field with the vector potential $A_k $, thus having the generalized
  momentum
\beq \label{yef__8_}
    P_{k} =p_{k} -\frac{e}{c} A_{k} , \qquad p_k =-i\hbar \nabla_k,
\eeq
  move in space with the quaternion metric being, up to the sign, the tensor
  part of the Q-multiplication rule (\ref{yef__3_}):
\beq \label{yef__9_}
    {\bf g}_{kn} \equiv -{\bf q}_{k} {\bf q}_{n}
        =\delta_{kn} -\eps_{knm} {\bf q}_{m} .
\eeq
  Then it is an easy exercise [7] to find that the Hamiltonian operator
  computed for this particle automatically comprises the initially heuristic
  Pauli spin term with a coefficient precisely coinciding with the Bohr
  magneton,
\bearr \label{yef__10_}
    H \equiv {\bf g}_{kn} P_{k} P_{n} \ \ \mapsto  \inch
\nnn \cm
    H=\frac{1}{2m} \left({\bf p}-\frac{e}{c} {\bf A}\right)^2
        -\frac{e\hbar}{2mc}{\bf B}\mbox{\boldmath $\sigma$};
\ear
  here the traditional 3D vector notations are used, $p_k \mapsto {\bf p}$,
  $A_{k} \mapsto {\bf A}$, ${\bf B}= {\rm curl}\, {\bf A}$, where $-i
  {\bf q}_{k} \mapsto \mbox{\boldmath $\sigma$}$ are the Pauli matrices.
  Introduction of the Pauli operator (\ref{yef__10_}) compels to replace the
  former ``scalar'' wave function of the electron with a 2D function-column,
  in fact a spinor.

\section{Complex numbers in a matrix form and their structure}

  A C-number
\beq \label{yef__11_}
    z=x+iy
\eeq
  can be regarded as a quaternion having nonzero real coefficients ($x,
  y\in {\mathbb R}$) at the real unit 1 and at only one of the vector
  units $i\to {\bf q}$,
\beq \label{yef__12_}
    z = x\cdot 1 + y\cdot {\bf q}.
\eeq
  One directly proves that each $2\times 2$ matrix of the generic form
\beq \label{yef__13_}
  {\bf q}=\frac{i}{\sqrt{T} } \left(\begin{array}{cc} {a} & {b} \\
                    {c} & {-a} \end{array}\right)
\eeq
  is an imaginary unit such that
\beq \label{yef__14_}
  {\bf q}^2  =- \left(\begin{array}{cc} {1} & {0} \\
                        {0} & {1} \end{array}\right)=-1
\eeq
  if
\beq \label{yef__15_}
    T\equiv a^2  +bc,
\eeq
  the components of the matrix (\ref{yef__13_}) being in general C-numbers
  (or functions) $a, b, c\in {\mathbb C}$, with the scalar ``interior
  imaginary unit'' $i$ involved only in the structure of ${\bf q}$. The
  unit (\ref{yef__13_}) is readily reduced to any of the vector units from
  the set (\ref{yef__7_}), and vice versa, it can be obtained from the
  vector units of the set (\ref{yef__7_}) using the transformations
  (\ref{yef__4_}) or (\ref{yef__6_}).

  Thus \eqs (\ref{yef__12_}), (\ref{yef__13_}) give a generic $2\times 2$
  matrix form of C-numbers obeying all laws of the C-algebra (the existence
  of summation, communicative an associative multiplication, division,
  conjugation, modulus, etc.). But unlike the traditional scalar
  description, a complex number in the form (\ref{yef__12_}) can be shown
  to have an elegant interior structure revealed in study of eigenfunctions
  of the matrix (\ref{yef__13_}) treated as an operator.

  The operator (\ref{yef__13_}) may have left (2D rows $\varphi$) and right
  (2D columns $\psi $) eigenfunctions (EF) satisfying the equations
\beq \label{yef__16_}
    \varphi   {\bf q}=k \varphi ,\qquad  {\bf q}\psi =l \psi,
\eeq
  with the eigenvalues $k,  l$. After simple algebra, one arrives at two
  sets of general solutions of \eqs (\ref{yef__16_}) $\varphi^+,\ \psi^+$
  and $\varphi^-, \psi^-$ attributed to positive
\beq \label{yef__17_}
        k^+ = l^+ =+ i,
\eeq
  and negative
\beq \label{yef__18_}
        k^- = l^- =-i ,
\eeq
  eigenvalues, respectively:\\
  for $a\ne T$:
\[
    \varphi^{\pm }=\frac{1}{\sqrt{2T}}
    \left(\sqrt{T\mp a},\ \mp \frac{b}{\sqrt{T\mp a} } \right),
\] $$
    \psi^{\pm } =\frac{1}{\sqrt{2T} } \left(\begin{array}{c}
    {\sqrt{T\mp a} } \\ {\mp \frac{c}{\dst \sqrt{T\mp a}}}
    \end{array}\right);                    \eqno (19a)
$$
  for $a=T,\ b=0$:
\[
    \varphi^+ =  (-c/2,\,1),\ \ \   \varphi^- =  (1,\,0),
\] $$
    \psi^+ = \left(\begin{array}{c} {0} \\ {1} \end{array}\right),
    \psi^- =   \left(\begin{array}{c} {1} \\ {c/2} \end{array}\right);
                        \eqno (19b)
$$
  for $a=T, c=0$:
\[
    \varphi^+ =  (0,\, 1),\ \ \ \varphi^- =  (1,\, b/2),
\] $$
    \psi^+ = \left(\begin{array}{c} {-b/2} \\ {1} \end{array}\right),
    \psi^- = \left(\begin{array}{c} {1} \\ {0} \end{array}\right).
                        \eqno (19c)
$$
  All EFs (19) possess the following properties. EFs of the same parity
  ($+$ or $-$) are normalized,
\setcounter{equation}{19}
\beq                           \label{yef__20_}
    \varphi^{\pm }  \psi^{\pm } =1,
\eeq
  EFs of opposite parities are automatically orthogonal:
\beq \label{yef__21_}
    \varphi^{\mp }  \psi^{\pm } = 0.
\eeq
  There is one more important feature: tensor products of EFs of the same
  parity give the two matrices
\beq \label{yef__22_}
    \psi^+  \varphi^+ =C^+ ,\ \ \ \psi^-  \varphi^- =C^-,
\eeq
  mutually orthogonal,
\beq \label{yef__23_}
        C^+ C^- =0,
\eeq
  and each having a zero determinant:
\beq \label{yef__24_}
        \det C^{\pm } =0;
\eeq
  one readily finds that any positive integer power $N$ of $C^{\pm }$
  returns the initial matrix:
\beq \label{yef__25_}
        C^{\pm N} =C^{\pm } ,
\eeq
  hence $C^{\pm}$ are idempotent matrices. The latter property makes it easy
  to find form \eqs (\ref{yef__16_}) that the units 1 and ${\bf q}$ are
  expressed in terms of the matrices $C^{\pm}$ as
\bearr \label{yef__26_}
    1 = C^+ +C^- =\psi^+ \varphi^+ + \psi^- \varphi^- ,
\yyy   \label{yef__27_}
    {\bf q} = i (C^+ -C^-) = i (\psi^+ \varphi^+ -\psi^- \varphi^-).
\ear
  \eqs (\ref{yef__26_}) and (\ref{yef__27_}) are fundamental mathematical
  equalities. They state that both units of the algebra of complex numbers,
  the real and imaginary ones, have an interior structure comprising more
  elementary mathematical objects. And since the imaginary unit ${\bf q}$ is
  one of the vector Q-units, these elementary objects, EFs, should be
  $SL(2,C)$, in particular $SU(\ref{yef__2_})$, spinors because the
  transformation (\ref{yef__6_}) providing stability of the Q-multiplication
  table (\ref{yef__3_}) is equivalent to transformations of the EFs only:
\beq \label{yef__28_}
    \psi'^{\pm } =U\psi^{\pm}, \ \ \ \varphi'^{\pm } =\varphi^{\pm} U^{-1}.
\eeq
  The spinor nature of the EFs will be a crucial point in our further
  discussion of a novel physics-geometry link, but let us now turn to
  the non-traditional image of C-numbers.

\section{The orthogonal form of C-numbers and the conic-gearing picture}

  Express the units in the C-number
\beq \label{yef__29_}
    z = x\cdot 1 + y\cdot  {\bf q}
\eeq
  in terms of the EFs using \eqs (\ref{yef__26_}) and (\ref{yef__27_}):
\bearr \label{yef__30_}
    z = x (\psi^+ \varphi^+ +\psi^- \varphi^- )+i y (\psi^+ \varphi^+
    -\psi^- \varphi^- )
\nnn \cm
    =(x +i y) \psi^+ \varphi^+ +(x-i y) \psi^- \varphi^-,
\ear
  or in terms of the idempotent matrices,
\beq \label{yef__31_}
    z = (x +i y) C^+ +(x-i y) C^- .
\eeq
  \eq (\ref{yef__31_}) represents the matrix C-number $z$ in an
  orthogonal form since a scalar C-number and its conjugate are
  coefficients of mutually orthogonal idempotents:
\beq \label{yef__32_}
        C^+ C^- =0.
\eeq
  The number (\ref{yef__31_}) is preferably rewritten in a polar format with
\bearr \label{yef__33_}
    r\equiv (x^2  +y^2  )^{1/2}, \ \ \  \alpha \equiv \arctan (y/x) ,
\yyy    \label{yef__34_}
    z = r \left(e^{i \alpha} C^+ +e^{-i \alpha} C^- \right).
\ear
  Any positive integer power of $z$ evidently preserves the orthogonal form.

  The bifurcation (\ref{yef__34_}) deserves a graphic image that will be
  step-by-step constructed as follows. The idempotents $C^+ ,\  C^- $ in
  a certain manner determine two orthogonal directions with attached
  components, each component, a scalar C-number, having its image on a
  complex plane. But \eq (\ref{yef__34_}) requires that there be two
  complex planes (two components) associated with the two orthogonal
  directions, hence the planes should be also conceived as being
  perpendicular to each other. Now temporarily fix the length $r $ of the
  components leaving the angle $\alpha $ variable; then the orthogonal
  component's planes are reduced to circles of radius $r$, and a change in
  the radius angle on one circle entails an angle change of the same value
  but of opposite sign on the orthogonal circle. Thus the components of the
  number (\ref{yef__34_}) behave as two mutually perpendicular discs
  touching at a point at their edges and able to rotate each other
  without slipping like a conic gear couple in a mechanical transmission
  (see the figure). It is obvious that a change of the discs' radius makes
  the picture enlarge or shrink without changing its shape, making a
  conformal transformation of the graph. This is the first ``direct''
  geometric image of the matrix C-number, but not the only one.

\begin{figure}
\center
\includegraphics[width=1\linewidth]{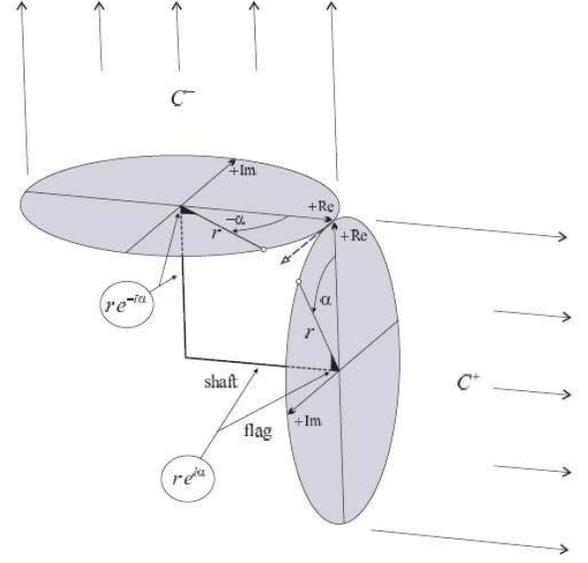}
\caption{Conic-gearing image of a complex number} \label{ris:image}
\end{figure}

  Supply the discs with ``shafts'', infinitesimally thin rods connecting the
  discs' centers with the central point of the graph where the rods
  intersect. Each rod obviously has a length equal to the disc radius $r$,
  and if at each rod a flag (``weathercock'') is attached marking the value
  of the angle $\alpha $ or -$\alpha $, then the ``rod \& flag'' graphs
  contain all information about the scalar C-number and its conjugate. A
  couple of orthogonal rods of certain (but equal) lengths twisted about
  themselves and having flags indicating the angles of the twist (angles of
  equal value but of opposite sign) is another, suggested here, geometric
  image of a C-number in the orthogonal matrix form. Both geometric images
  are shown in the conic gearing picture (see the figure), which in this
  case must be 3-dimensional. A question remains about depicting the objects
  $C^+$, $C^-$ being matrices, not vectors, nonetheless determining
  directions. In fact (it will be shown in the next section), the matrices
  are 2D geometric projectors onto a certain direction for any vector or
  tensor; this hints to portraying the two projectors as two orthogonal
  ``fluxes'' of arrows in 3D space.

\section{Geometry of spinors}

  Now it is useful to introduce two types of 2D indices. The first set of
  indices $A,B . . .=1,2$ is aimed at describing the components of
  $2\times 2$ matrices:
\beq \label{yef__35_}
    1\to \delta_{A}^{B},\ \ \
    {\bf q}\to q_{A  \cdot }^{\;B},\ \ \
    C  \to C_{A  \cdot }^{\;B};
\eeq
  the index position is important: the lower indices count columns, the
  upper ones enumerate rows. The left and right spinors acquire indices with
  different positions, which permits denoting them by the same (new) symbol
\beq \label{yef__36_}
    \varphi \to h_{A} ,\ \ \ \psi \to h^{A} .
\eeq
  Indices of the second type (in parentheses), also two-dimensional  $(K),
  (L) . . .=1,2$, replace the indicators of parity,
\bearr \nq \label{yef__37_}
    \varphi^- \to h_{(1)A},\ \  \psi^-  \to h_{(1)}^{B},\ \
    \varphi^+ \to h_{(2)A},\ \  \psi^+ \to h_{(2)}^{B},
\nnn
  {\rm or} \ \ \ \varphi^{\pm} \to h_{(K)A},\ \ \ \psi^{\pm} \to h_{(K)}^B,
\ear
  and
\beq \label{yef__38_}
    C^{\pm } \to C_{(K)A}{}^{\! B} ;
\eeq
  these indices will always be written at the lower position.  With these
  notations, the normalization and orthogonality equations (\ref{yef__20_}),
  (\ref{yef__21_}) are unified in the single equality (the Kronecker delta's
  indices are always used without parentheses)
\beq \label{yef__39_}
        h_{(K) A} h_{(L)}^{A} =\delta_{KL} ,
\eeq
  while \eq (\ref{yef__26_}) acquires the form
\beq \label{yef__40_}
        h_{(K) A} h_{(K)}^{B} =\delta_{A}^{B} .
\eeq
  \eqs (\ref{yef__39_}) and (\ref{yef__40_}) help us to make an important
  observation: they are identical to the conditions, well-known from
  differential geometry, for the generic Lam\'e coefficients $h_{(K)A}$,
  $h_{(K)}^{B}$ that locally link two coordinate systems,
\beq \label{yef__41_}
    y^{A} \leftrightarrow X_{(K)} ,
\eeq
  belonging to two different spaces, in our case 2D surfaces,
\beq \label{yef__42_}
    dX_{(K)} =h_{(K) A} dy^{A},\ \ \  dy^{B} =h_{(K)}^{B} dX_{(K)} .
\eeq
  The surfaces are tangent in their common point, and their common squared
  interval is
\bearr \label{yef__43_}
    dl^2  \equiv dy_{B} dy^{B} =h_{(K)B} h_{(L)}^{B} dX_{(K)} dX_{(L)}
\nnn \qquad
    =\delta_{KL} dX_{(K)} dX_{(L)} =dX_{(1)}^2  +dX_{(2)}^2.
\ear
  \eqs (\ref{yef__42_}) and (\ref{yef__43_}) state that one 2D space, having
  the coordinates $X_{(K)} $, possesses the Cartesian metric $\delta_{KL}$,
  so locally it is a plane, while the surface with the coordinates $y^{A} $
  has the metric
\beq \label{yef__44_}
    g_{AB} \equiv h_{(K)A} h_{(K)B}  ,
\eeq
  which is in fact \eq (\ref{yef__40_}) with both lowered indices. In
  general, the metric (\ref{yef__44_}) has variable components, and the
  surface it describes may be curved. According to the traditions of
  differential geometry, the 2D manifold with the metric (\ref{yef__44_})
  should be called a base space, and that with the Cartesian metric a
  tangent plane, while the set of Lam\'e coefficients $ h_{(K)}^{A} $
  linking the 2D spaces should be named a dyad (analogously to triads in
  the 3D case, tetrads in the 4D case etc.).

  Using the new notations, we write the idempotent matrices (\ref{yef__22_})
  with all lower indices:
\beq \label{yef__45_}
    C_{(1)AB} =h_{(1) A} h_{(1)B},\ \ \  C_{(2)AB} =h_{(2) A} h_{(2)B} ,
\eeq
  and find their components on the tangent plane; for this purpose, contract
  \eqs (\ref{yef__45_}) with the dyad $h_{(K)}^{A} $ using \eq
  (\ref{yef__39_}):
\bearr \label{yef__46_}
    C_{(1)(K)(L)} =h_{(1) A} h_{(1)B} h_{(K)}^{A} h_{(L)}^{B}
        =\delta_{1K} \delta_{1L} ,
\nnn
    C_{(2)(K)(L)} =\delta_{2K} \delta_{2L} .
\ear
  \eqs  (\ref{yef__46_}) directly show that the idempotent matrices are
  projectors onto one of the two orthogonal direction, there sum just
  yielding the component development of the Cartesian metric (Kronecker
  delta), the tangent equivalent of \eq (\ref{yef__26_})
\beq \label{yef__47_}
    \delta_{KL} =\delta_{1K} \delta_{1L} +\delta_{2K} \delta_{2L} ;
\eeq
  hence the Kronecker delta components $\delta_{1K}, \delta_{2K} $ are the
  simplest (plane) spinors forming the tangent plane. The imaginary unit
  (\ref{yef__27_}) written in terms of the plane spinors has the form of
  the imaginary Lorentz-type metric
\beq \label{yef__48_}
    q_{KL} =-i (\delta_{1K} \delta_{1L} -\delta_{2K} \delta_{2L} ),
\eeq
  its components coinciding with those of the canonical Q-unit ${\bf q}_{3}$
  form, \eq (\ref{yef__7_}). The plane spinors $\delta_{1K},  \delta_{2K} $
  are obviously eigenfunctions of the vector operator (\ref{yef__27_}), but
  they can be used as well to build the components of any other Q-unit from
  \eq (\ref{yef__7_}), e.g., ${\bf q}_{2} $:
\beq \label{yef__49_}
    {\bf q}_{2} \mapsto   q_{LN}
    =\delta_{1N} \delta_{2L} -\delta_{2N} \delta_{1L}.
\eeq
  A product of \eqs (\ref{yef__49_}) and (\ref{yef__48_}) gives an
  expression of ${\bf q}_1$ from \eq (\ref{yef__7_}) in terms of the plane
  spinors:
\bearr
    {\bf q}_{1} ={\bf q}_{2} {\bf q}_{3} \ \mapsto  \
    q_{KN} =q_{KL} q_{LN}
\nnn        \label{yef__50_}
    =(\delta_{1N} \delta_{2L} -\delta_{2N} \delta_{1L} )
    (-i)(\delta_{1K} \delta_{1L} -\delta_{2K} \delta_{2L} )
\nnn
    =i (\delta_{1K} \delta_{2N} +\delta_{2K} \delta_{1N} ).
\ear
  \eqs (\ref{yef__48_})--(\ref{yef__50_}) show that the plane spinors
  belonging to some tangent plane (in this case as Lam\'e coefficients
  linking the Cartesian systems of two planes coordinates) are at the same
  time elements for construction of all canonical vector Q-units
  (\ref{yef__7_}). This establishes a direct --- and quite technological ---
  mathematical relationship between an abstract plane (a screen) and each
  direction of the 3D space (the physical space).

\section{Concluding remarks}

  There are three main results of the study to be emphasized. First, the
  quaternion space, a natural geometric consequence of Q-algebra, gives rise
  to the Pauli term in the quantum-mechanical Hamiltonian of the electron.
  Second, the spinors, an extension of the wave function, are revealed to be
  structural elements of Q-units, which is a pure mathematical fact
  suggesting a novel geometric image of C-numbers, a dimensional reduction
  of quaternions. And the third significant issue of the study is
  establishing of a geometric meaning of the set of spinor functions, basic
  elements of 3D vector Q-units, as Lam\'e coefficients (dyads)
  interconnecting two mutually tangent surfaces. Investigations of variable
  (non-planar) spinors are expected to reveal the mutual dependence of the
  curved spinor-born base surface and the respective dynamic points of the
  3D (physical) space endowed with a vector Q-triad. An intriguing subject
  is an analysis of limits on the curvature and/or regularity of the
  local surface that can result in discontinuity and spectrum of the dyad
  function thus possibly leading to a better understanding of quantum
  phenomena. Further explorations of fundamental structures of H-numbers
  will hopefully allow extending the knowledge of this surprisingly rich
  ``math-medium'' and its links with physics.

\end{document}